\documentclass[10pt,a4paper]{article}
\usepackage{amsfonts}
\usepackage{amsthm}
\usepackage{amsmath}
\usepackage{latexsym}
\usepackage{amssymb}
\usepackage{amstext}
\usepackage{fancyhdr}
\usepackage{indentfirst}
\usepackage{graphicx}
\usepackage{epsfig}
\usepackage{pgf}
\usepackage{listings}
\usepackage{cases}
\usepackage{epstopdf}
\usepackage{graphicx}
\usepackage{enumerate}
\date{}

\begin{document}
\title{Signless Laplacian State Transfer on Vertex Complemented Coronae}
\author{Ke-Yu Zhu$^a$, Gui-Xian Tian$^a$\footnote{Corresponding author. E-mail address: gxtian@zjnu.cn or guixiantian@163.com.}, Shu-Yu Cui$^{b}$\\
    {\small{\it $^a$School of Mathematical Sciences,}}
	{\small{\it Zhejiang Normal University, Jinhua, Zhejiang, 321004, P.R. China}}\\
	{\small{\it $^b$Xingzhi College, Zhejiang Normal University, Jinhua, Zhejiang, 321004, P.R. China}}}
\maketitle

\begin{abstract}
Given a graph $G$ with vertex set $V(G)=\{v_1,v_2,\ldots,v_{n_1}\}$ and a graph $H$ of order $n_2$, the vertex complemented corona, denoted by $G\tilde{\circ}{H}$, is the graph produced by copying $H$ $n_1$ times, with the $i$-th copy of $H$ corresponding to the vertex $v_i$, and then adding edges between any vertex in $V(G)\setminus\{v_{i}\}$ and any vertex of the $i$-th copy of $H$. The present article deals with quantum state transfer of vertex complemented coronae concerning signless Laplacian matrix. Our research investigates conditions in which signless Laplacian perfect state transfer exists or not on vertex complemented coronae. Additionally, we also provide some mild conditions for the class of graphs under consideration that allow signless Laplacian pretty good state transfer.

\emph{AMS classification:} 05C50; 15A18; 81P45

\emph{Keywords:} Quantum state transfer; Periodic; Signless Laplacian; Vertex complemented corona
\end{abstract}

\section{Introduction}

Quantum state transfer in quantum networks is an essential topic in the research of quantum communication protocols. The key issue is how to avoid losing any or very little information in the process of transmission. Farhi and Gutmann \cite{Farhi1998} developed a quantum-mechanical algorithm to evolve a state across a tree, starting at the root. In their research, for a simple connected graph $G$ with vertex set $V(G)$, the unitary time evolution operator of continuous-time quantum walk on $G$ relative to a Hamiltonian $M$ is defined as $U_M(\tau)=\exp(-\mathrm{i}\tau M)$, note that $\mathrm{i}^2=-1$. In 2003, Bose \cite{Bose03} showed that short-distance quantum communications are carried out using an unmodulated and unmeasured spin chain. Next year, Christandl et al. \cite{Christandl04} focused on information communications in quantum spin chains, and they originally put forward the notion of perfect state transfer (popularly shortened to PST). If and only if $\left|\exp(-\mathrm{i}\tau M)_{uv}\right|^2=1$, where $u, v$ are two vertices in $G$, we call $G$ allows PST between these two vertices at time $\tau$. Additionally, $v$ is considered to be periodic at time $\tau$ if $u$ equals to $v$.

In 2012, Godsil \cite{Godsil12} described the situations in which PST happens, it was proved that, given any positive integer $k$, the set of graphs allowing PST is restricted to those with maximum degree $k$. Nonetheless, it is challenging to identify a graph that permits PST since PST is quite uncommon. Thus, it's usually more beneficial to concerning the subsequent relaxation of PST, viz., pretty good state transfer (PGST) which is attributed to Godsil \cite{Godsil11}. We call $G$ exhibits PGST if and only if at some time $\tau$, the inequality $|\exp(-\mathrm{i}\tau M)_{uv}|^2\geq1-\epsilon$ holds for arbitrary $\epsilon>0$.

Quantum state transfer has aroused the interest of numerous academics and has been applied extensively. Noteworthy, quantum state transfer about adjacency and Laplacian matrice are the main subject of current literature. This paper primarily focuses on quantum state transfer in a variant of corona product operation of graphs. Now we only recall some observations concerning the state transfer on corona operation of graphs. In 2017, Ackelsberg et al. \cite{Ackelsberg17} evaluated the conditions in which corona graph $G\circ H$ displays PST and PGST relative to adjacency matrix, and based on this, they constructed some family of graphs exhibiting PST and PGST. Almost simultaneously, they also studied PST and PGST on corona graphs relative to Laplacian matrix. It was proved \cite{Ackelsberg16} that $G\circ H$ avoids Laplacian-PST whenever $G$ has more than one vertex, but if the conditions are relaxed in some way, then $G\circ H$ exhibits Laplacian-PGST. In 2022, the research about PST (resp., Laplacian-PST) on vertex complemented coronae has been conducted in \cite{Wang22} (resp., \cite{Wang22+}) and some suitable requirements for the graphs to exhibit PGST or Laplacian-PGST were also established. Some results about PST and PGST on neighborhood coronae can be found in \cite{Zhang23}. Laplacian-PST and Laplacian-PGST on other variant of coronae, such as extended neighborhood coronae, edge coronae and edge complemented coronae, have been intensively studied in \cite{Li21,Li22,Wang21}.

In contrast, there are only a few results about signless Laplacian state transfer. In 2016, Alvir et al. \cite{Alvir16} explored the existence of signless-Laplacian-PST on graphs. They established an elegant relationship between signless-Laplacian-PST of a graph $G$ and PST of its line graph $l(G)$. It has also been observed that signless-Laplacian-PST and PST are equivalent for regular graphs. These open a new perspective in the study of PST on graphs and will help us to find more graphs with PST. In 2021, Tian et al. \cite{Tian21} studied the PST and PGST relative to signless Laplacian matrix on coronae. It is proved that several special coronae lack signless-Laplacian-PST, such as $\overline{nK_2} \circ H$ for any regular graph $H$, $K_2 \circ H$ for an even-order regular graph $H$ and so on. However, $\overline{nK_2} \circ K_1$ for $n>2$ and $K_2 \circ H$ for any regular graph $H$ allow signless-Laplacian-PGST. Resently, PST and PGST relative to signless Laplacian matrix on $\mathcal{Q}$-graphs were also investigated in \cite{Zhang22}.

Inspired by the research above, we focus on whether PST and PGST relative to signless Laplacian matrix occur on vertex complemented coronae, whose concept is attributed to Gayathri \cite{Gayathri20}. Given a graph $G$ and its vertex set $V(G)=\{v_1,\ldots,v_{n_1}\}$, let $H$ be a graph with $n_2$ vertices. The vertex complemented corona $G\tilde{\circ}{H}$ is the graph produced by copying $H$ $n_1$ times, with the $i$-th copy of $H$ corresponding to the vertex $v_i$, and then adding edges between each vertex in $V(G)\setminus\{v_{i}\}$ and each vertex of the $i$-th copy of $H$ ($i=1,2,\ldots,n_1$). First we characterize the signless Laplacian eigenvalues and eigenvectors of $G\tilde{\circ}{H}$, which are applied to provide signless Laplacian spectral decomposition of vertex complemented corona. Then, by analyzing periodic conditions, we obtain several special vertex complemented coronae with no signless-Laplacian-PST. In particular, we state that if $H$ is a regular graph of order $k$, after which $K_2\tilde{\circ}{H}$ avoids signless-Laplacian-PST for $k=1$ or any prime number $k$. Observe that $K_2\tilde{\circ}{H}$ and $K_2{\circ}{H}$ are the same. In \cite{Tian21}, Tian et al. proved that $K_2{\circ}{H}$ avoids signless-Laplacian-PST for any even number $k$. So, we suspect that the conclusion is always valid for any $k$. We also prove that, for a regular graph $G$, the vertex complemented corona $G\tilde{\circ}{\overline{K_m}}$ allows signless-Laplacian-PGST under some particular circumstances. In contrast, if $G$ is some special distance-regular graph, then  $G\tilde{\circ}{\overline{K_1}}$ has no signless-Laplacian-PST (see Example 2). Finally, we prove that the graph $\overline{nK_2} \tilde{\circ} K_1$ allows signless-Laplacian-PGST in nearly all situations. What's interesting is that this result also holds for $\overline{nK_2} \circ K_1$ (see \cite{Tian21}).

\section{Preliminaries}

Given a connected graph $G$, let $A(G)$ and $D(G)$ denote its adjacency matrix and degree diagonal matrix. We use $Q(G)=D(G)+A(G)$ to represent the signless Laplacian matrix of $G$, and the collection $\theta_0>\theta_1>\cdots>\theta_m$ represent all distinct eigenvalues of $Q(G)$, which is referred to as $\text {Spec}_Q(G)$. The spectral decomposition theorem allows us to write
\begin{equation*}
Q(G)=\sum_{r=0}^m\theta_r F_{\theta_r}(G),
\end{equation*}
where $F_{\theta_r}(G)$ denotes the signless Laplacian eigenprojector relative to the eigenvalue $\theta_r$ of $Q(G)$. Since all eigenprojectors are orthogonal, then the transition matrix $U_{Q(G)}(\tau )$ is provided by
\begin{equation*}
U_{Q(G)}(\tau)=\exp(-\mathrm{i}\tau Q(G))=\sum_{r=0}^m\exp(-\mathrm{i}\tau \theta_r)F_{\theta_r}.
\end{equation*}
The signless Laplacian eigenvalue support of a vertex $u$ in $G$, portrayed as $\mathrm{supp}_G(u)$, is the set of eigenvalues $\theta$ of $Q(G)$ which ensures $F_\theta(G)\mathbf{e}_u\neq0$, where $\mathbf e_u$ is the characteristic vector of the vertex $u$. We say two vertices $u$ and $v$ are signless Laplacian strongly cospectral if and only if $F_\theta(G)\mathbf{e}_u=\pm F_\theta(G)\mathbf{e}_v$, for $\forall\; \theta \in \text {Spec}_Q(G)$. Based on this, we define $\Lambda^+$ for the set of $\theta\in \mathrm{supp}_G(u)$ such that $F_{\lambda}(G)\mathbf{e}_{u}=F_{\lambda}(G)\mathbf{e}_{v}$, and $\Lambda^-$ for the set of $\theta\in \mathrm{supp}_G(u)$ such that $F_{\lambda}(G)\mathbf{e}_{u}=-F_{\lambda}(G)\mathbf{e}_{v}$.\\

The following results, due to Coutinho \cite{Coutinho14}, are beneficial to us.

\paragraph{Theorem 2.1.}(Coutinho \cite{Coutinho14}) If $G$ allows PST from $u$ to $v$ at time
$\tau$, then $G$ is periodic at both $u$ and $v$ at time $2\tau $.

\paragraph{Theorem 2.2.}(Coutinho \cite{Coutinho14}) Let $u$ and $v$ be two vertices in $G$. Also let $S=\mathrm{supp}_{G}(u)$ and $\theta_0>\theta_1>\cdots>\theta_k$ be the signless Laplacian eigenvalues in $S$. Then graph $G$ allows signless-Laplacian-PST from $u$ to $v$ if and only if each of the subsequent requirements is met.
\begin{enumerate}[a)]
\item $u$ and $v$ are signless Laplacian strongly cospectral.
\item Either every element of $\mathrm{supp}_G(u)$ is an integer, or a quadratic integer. Furthermore, for $r\in \{0,...,k\}$, there is a square-free integer $\Delta$ that satisfies
\begin{equation*}
\theta_r=\frac12(a+b_r\sqrt{\Delta}),
\end{equation*}
where $a,b_0,...,b_k$ are all integers.
\item Set $g=\gcd\left(\left\{\frac{\theta_0-\theta_r}{\sqrt{\Delta}}\right\}_{r=0}^k\right)$. Then
\begin{enumerate}[i)]
\item $\theta_r\in\Lambda^+$ if and only if $\frac{\theta_0-\theta_r}{g\sqrt{\Delta}}$ is even;
\item $\theta_r\in\Lambda^-$ if and only if $\frac{\theta_0-\theta_r}{g\sqrt{\Delta}}$ is odd.
\end{enumerate}
\end{enumerate}
If each of the requirements above holds and there exists signless-Laplacian-PST from $u$ to $v$ at time $\tau $, then the minimum time $\tau_0=\frac\pi{g\sqrt{\Delta}}$ with the corresponding phase $\lambda=\mathrm{e}^{i\tau \theta_{0}}$.

Moreover, if $u=v$, we say $G$ is periodic at $u$ if and only if every element in $\mathrm{supp}_G(u)$ is either an integer, or a quadratic integer.

\paragraph{Theorem 2.3.}(Hardy and Wright \cite{Hardy00}) Suppose that $1,\lambda_1,\ldots,\lambda_m$ are linearly independent on $\mathbb{Q}$. Following this, for arbitrary numbers $\alpha_1,\ldots,\alpha_m\in\mathbb{R}$  and $\mathbf{\epsilon}\in\mathbb{R^+}$, there must exist integers $l$ and $q_1,\ldots,q_m$ of which satisfy
\begin{equation}\label{1}
|l\lambda_k-\alpha_k-q_k|<\epsilon,\;\; \forall \;k\in \{1,...,m\}.
\end{equation}

For the purpose of simplicity, we write $\alpha\approx\beta $ rather than $|\alpha-\beta|<\epsilon $ when $\epsilon$ is arbitrarily small. Thereby, (\ref{1}) can be written as $l\lambda_k-q_k\approx\alpha_k$.\\

Using the theorem below, we can determine the linearly independence of some set on $\mathbb{Q}$.

\paragraph{Theorem 2.4.}(Richards \cite{Richards1974}) The set $\{\sqrt{\Delta}:\Delta \; \text{is a square-free integer}\}$ is linearly independent on $\mathbb{Q}$.\\

Given a distance-regular graph $G$ with diameter $d$, let $G_d$ be the graph with two vertices adjacent if the distance between them is $d$. The adjacency matrix of $G_d$ is shown by $A_d(G)$. What below is attributed to Coutinho et al. \cite{Coutinho15}.

\paragraph{Theorem 2.5.}(Coutinho et al. \cite{Coutinho15}) Suppose $G$ is an antipodal distance-regular graph with classes of size two and all distinct eigenvalues $\theta_0>\theta_1>\cdots>\theta_d$ with respective eigenprojectors $F_{\theta_0}(G),F_{\theta_1}(G),\ldots,F_{\theta_d}(G)$. Then,
\begin{equation*}
A_d(G)F_{\theta_i}(G)=(-1)^iF_{\theta_r}(G),\;\; \forall \;i\in \{0,1,...,d\}.
\end{equation*}

\section{Signless Laplacian of vertex complemented coronae}

Let $G$ be a graph with vertex set $V(G)=\{v_1,...,v_{n_1}\}$ and $H$ a graph of order $n_2$. To make things easier, the vertex set of $G\tilde{\circ}{H}$ is indicated in the following:

\begin{equation*}
V(G\tilde{\circ}{H})=V(G)\times\left(\{0\}\cup V(H)\right),
\end{equation*}
and the edge set satisfies the relation as follows:
\begin{equation*}
(v_i,w)\sim(v_j,w')\Longleftrightarrow\left\{\begin{array}{ll}w=w'=0\text{ and }v_i\sim v_j\text{ in }G,\;\text{or}\\v_i=v_j\text{ and }w\sim w'\text{ in }H,\;\text{or}\\v_i\neq v_j\text{and precisely one of }w\text{ and }w'\text{ is }0.\end{array}\right.
\end{equation*}

If $G$ and $H$ are both regular graphs, the signless Laplacian eigenvalues and eigenprojectors of $G\tilde{\circ}{H}$ are determined as below.

\paragraph{Theorem 3.1.} Suppose $G$ is an $r_1$-regular connected graph with $n_1$ vertices and $H$ is an $r_2$-regular graph with $n_2$ vertices. Then the signless Laplacian eigenvalues and eigenprojectors of $G\tilde{\circ}{H}$ are provided by:
\begin{enumerate}[(a)]
\item For each signless Laplacian eigenvalue $\mu$ of $H$, $n_1-1+\mu$ is a signless Laplacian eigenvalue of $G\tilde{\circ}{H}$ with the eigenprojector
\begin{equation}\label{2}
F_{n_1-1+\mu}=\left(\begin{array}{cc}\mathbf{0}&\mathbf{0}\\\mathbf{0}&I_{n_1}\otimes \left(F_\mu(H)-\delta_{\mu,2r_2}\cdot\frac{1}{n_2}J_{n_2}\right)\end{array}\right),
\end{equation}
where $F_\mu(H)$ stands for the eigenprojector corresponding to the signless Laplacian eigenvalue $\mu$ of $H$, and the function $\delta_{\mu,2r_2}$ has the form, as below
\begin{equation*}
\delta_{\mu,2r_2}=\left\{\begin{array}{cc}1,&\mu=2r_2,\\0,&\mu\neq 2r_2.\end{array}\right.
\end{equation*}
\item For each signless Laplacian eigenvalue $\theta$ of $G$ except that the eigenvalue $2r_1$, $\theta_{\pm}=\frac{1}{2}(\theta+s+t\pm\sqrt{(\theta-s+t)^{2}+4n_{2}})$ are signless Laplacian eigenvalues of $G\tilde{\circ}{H}$ with the eigenprojectors
\begin{equation}\label{3}
\begin{aligned}
\left.F_{\theta_\pm}(G\tilde{\circ}{H})=F_\theta(G)\otimes\frac1{(s-\theta_\pm)^2+n_2}\left(\begin{array}{cc}(s-\theta_\pm)^2&(s-\theta_\pm)\mathbf{j}_{n_2}^T\\(s-\theta_\pm)\mathbf{j}_{n_2}&J_{n_2}\end{array}\right.\right),
\end{aligned}
\end{equation}
where $s=n_1+2r_2-1$ and $t=n_2(n_1-1)$.
\item $r_{\pm}=\frac{1}{2}(2r_{1}+s+t)\pm\sqrt{(2r_{1}-s+t)^{2}+4n_2(n_1-1)^{2}}$ are signless Laplacian eigenvalues of $G\tilde{\circ}{H}$ with the eigenprojectors
\begin{equation}\small\label{4}
\begin{aligned}
F_{r_\pm}(G\tilde{\circ}{H})=F_{2r_1}(G)\otimes\frac1{(s-r_\pm)^2+n_2(1-n_1)^2}
\left(\begin{array}{cc}(s-r_\pm)^2&(s-r_\pm)(1-n_1)\mathbf{j}_{n_2}^T\\(s-r_\pm)(1-n_1)\mathbf{j}_{n_2}&(1-n_1)^2J_{n_2}\end{array}\right).
\end{aligned}
\end{equation}
\end{enumerate}
Moreover, the spectral decomposition of $Q(G\tilde{\circ}{H})$ is provided by:
\begin{equation*}
Q(G\tilde{\circ}{H})=\left(\sum_{\theta\in\mathrm{Spec}_Q(G)\backslash\{2r_1\}}\sum_{\pm}\theta_\pm F_{\theta_\pm}\right)+\sum_{\mu\in\mathrm{Spec}_Q(H)}(n_1-1+\mu) F_\mu+\sum_{\pm}r_\pm F_{r_\pm}.
\end{equation*}

\begin{proof}
Suppose that $\mu$ is a signless Laplacian eigenvalue of $H$, define $B_\mu$ as the orthonormal basis corresponding to the eigenspace of $\mu$, of which orthogonal to $\mathbf{j}_{n_2}$. It's important to note that $B_{2r_2}$ is always empty whenever $H$ is connected. And if $H$ is a disconnected graph, there are other eigenvectors, orthogonal to $\mathbf{j}_{n_2}$, of $Q(H)$ corresponding to $2r_2$. Assume that $\mu$ has multiplicity $m$, thus the cardinality of $B_\mu$ is $m-1$ when $\mu=2r_2$, and $m$ otherwise. Note that the signless Laplacian matrix $Q(G\tilde{\circ}{H})$ is the following:
\begin{equation*}
Q(G\tilde{\circ}{H})=\left(\begin{array}{cc}Q(G)+(n_1-1)n_2I_{n_1}&M\otimes\mathbf{j}_{n_2}^T\\M^T\otimes\mathbf{j}_{n_2}&I_{n_1}\otimes(Q(H) +(n_1-1)I_{n_2})\end{array}\right)
\end{equation*}
with $M=J_{n_1}-I_{n_1}$. From this fact, we can get that, for each $\mathbf x\in B_\mu$,
 \begin{equation*}
Q(G\tilde{\circ}H)\mathbf{e}_{u}\otimes\left[{\begin{array}{c}0\\\mathbf{x}\end{array}}\right]=(n_1-1+\mu)\mathbf{e}_{u}\otimes\left[{\begin{array}{c}0\\\mathbf{x}\end{array}}\right].
 \end{equation*}
It follows that the orthonormal basis corresponding to the eigenspace of $n_1-1+\mu$ in $G\tilde{\circ}{H}$ has the form:
\begin{equation*}
\left\{\mathbf{e}_{u}\otimes\left[\begin{array}{c}0\\\mathbf{x}\end{array}\right]:u\in V(G),\;\mathbf{x}\in B_{\mu}\right\}.
\end{equation*}
From the above argument, we find that the cardinality of this orthonormal basis is $n_1(m-1)$ when $\mu=2r_2$, and $n_1m$ otherwise. It is apparent that the signless Laplacian eigenprojector of $\mu$ in $H$ has the form
\begin{equation*}
F_\mu(H)=\begin{cases}\sum_{\mathbf{x}\in B_\mu}\mathbf{x}\mathbf{x}^T+\frac1n_2J_{n_2}&\mathrm{if~}\mu=2r_2,\\\sum_{\mathbf{x}\in B_\mu}\mathbf{x}\mathbf{x}^T&\mathrm{otherwise}.\end{cases}
\end{equation*}
On the basis of this fact, we find that $n_1-1+\mu$ is a signless Laplacian eigenvales of $G\tilde{\circ}{H}$ with the eigenprojector in the form of (\ref{2}).

In the next step, we take $\theta_{\pm}=\frac{1}{2}(\theta+s+t\pm\sqrt{(\theta-s+t))^{2}+4n_{2}})$ for each signless Laplacian eigenvalue $\theta$ of $G$ ($\theta\neq 2r_1$). Assume that $\mathbf{y}$ is an associated signless Laplacian eigenvector of $\theta$, notice that $\mathbf{y}\bot \mathbf{j}_{n_1}$. By a simple calculation, one has
\begin{equation*}
Q(G\tilde{\circ}{H})\left(\mathbf{y}\otimes\begin{bmatrix}1\\\frac{1}{s-\theta_\pm}\mathbf{j}_{n_2}\end{bmatrix}\right)=\theta_\pm\left(\mathbf{y}\otimes\begin{bmatrix}1\\\frac{1}{s-\theta_\pm}\mathbf{j}_{n_2}\end{bmatrix}\right).
\end{equation*}

In the same way, we define $r_{\pm}=\frac{1}{2}(2r_1s+t)\pm\sqrt{(2r_1-s+t))^{2}+4n_2(n_1-1)^{2}}$, for the signless Laplacian eigenvalue $2r_1$ of $G$ and its associated signless Laplacian eigenvector $\mathbf{z}$. Notice that $Q(G)\mathbf{z}=2r_1\mathbf{z}$, $\mathbf{z}=\frac{1}{\sqrt{n_{1}}}\mathbf{j}_{n_{1}}$ and $M\mathbf{z}=(n_1-1)\mathbf{z}$, one obtains
\begin{equation*}
Q(G\tilde{\circ}{H})\left(\mathbf{z}\otimes\begin{bmatrix}1\\\frac{n_1-1}{r_\pm-s}\mathbf{j}_{n_2}\end{bmatrix}\right)=r_\pm\left(\mathbf{z}\otimes\begin{bmatrix}1\\\frac{n_1-1}{r_\pm-s}\mathbf{j}_{n_2}\end{bmatrix}\right).
\end{equation*}

Normalizing these signless Laplacian eigenvectors above yields the eigenprojectors of $\theta_\pm$ and $r_\pm$ as follows:
\begin{equation*}
\left.F_{\theta_\pm}(G\tilde{\circ}{H})=F_\theta(G)\otimes\frac1{(s-\theta_\pm)^2+n_2}\left(\begin{array}{cc}(s-\theta_\pm)^2&(s-\theta_\pm)\mathbf{j}_{n_2}^T\\(s-\theta_\pm)\mathbf{j}_{n_2}&J_{n_2}\end{array}\right.\right)
\end{equation*}
and
\begin{equation*}
\begin{aligned}
F_{r_\pm}(G\tilde{\circ}{H})=F_{2r_1}(G)\otimes\frac1{(s-r_\pm)^2+n_2(1-n_1)^2}
\left(\begin{array}{cc}(s-r_\pm)^2&(s-r_\pm)(1-n_1)\mathbf{j}_{n_2}^T\\(s-r_\pm)(1-n_1)\mathbf{j}_{n_2}&(1-n_1)^2J_{n_2}\end{array}\right).
\end{aligned}
\end{equation*}
Finally, the signless Laplacian spectral decomposition of $G\tilde{\circ}{H}$ can be obtained directly from what has been discussed above. This concludes our proof.
\end{proof}

Based on Theorem 3.1 and by a tedious calculation, we derive the explicit expression for the elements in $U_{Q(G\tilde{\circ}{H})}(\tau)$, which will be frequently utilized to analyze the existence of state transfer based on signless Laplacian on $G\tilde{\circ}{H}$.

\paragraph{Theorem 3.2.} Suppose the graphs $G$ and $H$ are defined as Theorem 3.1, and $u,v\in V(G)$. Then
\begin{equation}\small\label{5}
\begin{aligned}
\mathbf{e}_{(u,0)}e^{-\mathrm{i}\tau Q(G\tilde{\circ}{H})}\mathbf{e}_{(v,0)} =&\sum_{\theta\in\mathrm{Spec}_{Q}(G)\setminus\{2r_1\}}e^{-\mathrm{i}\tau \frac{\theta+s+t}{2}}
\left(\cos\left(\frac{\Lambda_{\theta}\tau}{2}\right)-\mathrm{i}\frac{\theta-s+t}{\Lambda_{\theta}}\sin\left(\frac{\Lambda_{\theta}\tau}{2}\right)\right)\mathbf{e}_{u}^TF_{\theta}(G)\mathbf{e}_{v} \\
&+e^{-\mathrm{i}\tau \frac{2r_{1}+s+t}{2}}\left(\cos\left(\frac{\Lambda_{r}\tau}{2}\right)-\mathrm{i}\frac{2r_1-s+t}{\Lambda_{r}}\sin\left(\frac{\Lambda_{r}\tau}{2}\right)\right)\mathbf{e}_{u}^TF_{2r_1}(G)\mathbf{e}_{v},
\end{aligned}
\end{equation}
where $ \Lambda_{\theta} = \sqrt{(\theta-s+t)^{2}+4n_2}$ and $\Lambda_r=\sqrt{(2r_1-s+t)^2+4n_2(n_1-1)^2}$.
\begin{proof} Note that $\theta_\pm=\frac{1}{2}(\theta+s+t\pm\Lambda_\theta)$ ($\theta\neq 2r_1$), $r_\pm=\frac{1}{2}(2r_1+s+t\pm\Lambda_r)$. By Theorem 3.1, the transition element from $(u,0)$ to $(v,0)$ has the following form:
\begin{equation}\small\label{6}
\begin{aligned}
\mathbf{e}_{(u,0)}^{T} \mathrm{e}^{-\mathrm{i}\tau Q(G\tilde{\circ}H)}\mathbf{e}_{(v,0)}=&\sum_{\theta\in\mathrm{Spec}_{Q}(G)\setminus\{2r_1\}}e^{-\mathrm{i}\tau \frac{\theta+s+t}{2}}\mathbf{e}_u^TF_\theta(G)\mathbf{e}_v\left(\sum_{\pm}e^{\mp \mathrm{i}\tau \frac{\Lambda_{\theta}\tau}{2}}\frac{(s-\theta_\pm)^2}{(s-\theta_\pm)^2+n_2}\right)\\
&+e^{-\mathrm{i}\tau \frac{2r_1+s+t}{2}}\mathbf{e}_u^T F_{2r_1}(G)\mathbf{e}_v\left(\sum_{\pm}e^{\mp\mathrm{i}\tau \frac{\Lambda_r}{2}}\frac{(s-r_\pm)^2}{(s-r_\pm)^2+n_2(1-n_1)^2}\right).
\end{aligned}
\end{equation}
Due to
\begin{equation}\label{7}
\prod_\pm(s-\theta_\pm)=-n_2,\quad\prod_\pm(s-r_\pm)=-n_2(1-n_1)^2.
\end{equation}
It follows that,
\begin{equation}\label{8}
\prod_\pm\left((s-\theta_\pm)^2+n_2\right)=n_2\Lambda_\theta^2,\quad \prod_\pm\left((s-r_\pm)^2+n_2(1-n_1)^2\right)=n_2(1-n_1)^2\Lambda_r^2.
\end{equation}
Now, using (\ref{7}), (\ref{8}) and by a simple calculation, one obtains
\begin{equation*}
\sum_{\pm}e^{\mp\mathrm{i}\tau \frac{\Lambda_{\theta}}{2}}\frac{(s-\theta_\pm)^2}{(s-\theta_\pm)^2+n_2}=\cos\left(\frac{\Lambda_{\theta}\tau}{2}\right)-\mathrm{i}\frac{\theta-s+t}{\Lambda_{\theta}}\sin\left(\frac{\Lambda_{\theta}\tau}{2}\right)
\end{equation*}
and
\begin{equation*}
\sum_{\pm}e^{\mp\mathrm{i}\tau \frac{\Lambda_r}{2}}\frac{(s-r_\pm)^2}{(s-r_\pm)^2+n_2(1-n_1)^2}=\cos\left(\frac{\Lambda_r\tau}{2}\right)-\mathrm{i}\frac{2r_1-s+t}{\Lambda_r}\sin\left(\frac{\Lambda_r\tau}{2}\right).
\end{equation*}
Finally, plugging these formulas into (\ref{6}), the desired result (\ref{5}) follows.
\end{proof}

\section{Signless-Laplacian-PST}

Theorem 2.1 indicates the relationship between periodic vertices and PST, that is, for PST to be possible, the two vertices must be periodic. From this fact, we shall attempt to find whether there exists signless-Laplacian-PST on vertex complemented coronae through periodicity of vertices.

\paragraph{Lemma 4.1.} Suppose that the graphs $G$ and $H$ are as stated in Theorem 3.1. Then, for $u\in V(G)$ and $v\in V(H)$, $(u,0)$ is periodic on $G\tilde{\circ}{H}$ whenever $(u,v)$ is periodic on $G\tilde{\circ}{H}$.

\begin{proof} The eigenprojectors provided in Theorem 3.1 implies that $\mathrm{supp}_{G\tilde{\circ}H}(u,0)\subseteq\mathrm{supp}_{G\tilde{\circ}H}(u,v)$, which concludes this result.
\end{proof}

Next, we demonstrate that periodicity allows us to impose tight constraints on signless Laplacian eigenvalues. What below states some sufficient conditions to have periodic vertices. We further find that most vertex complemented coronae avoid signless-Laplacian-PST by proving the existence of periodic vertices.

\paragraph{Theorem 4.2.} Suppose that the graphs $G$ and $H$ are as stated in Theorem 3.1.
\begin{enumerate}[(i)]
\item If $2r_1+t\neq s$, then $(v,0)$ is periodic in $G\tilde{\circ}{H}$ if, and only if $\theta-s+t$, $\sqrt{(\theta-s+t)^2+4n_2}$ and $\sqrt{(2r_1-s+t)^2+4n_2(n_1-1)^2}$ are all integers for any signless Laplacian eigenvalue $\theta\in\mathrm{supp}_G(v)\setminus\{2r_1\}$.
\item If $2r_1+t=s$, then $(v,0)$ is periodic in $G\tilde{\circ}{H}$ if, and only if there is a square-free $\Delta\in\mathbb{Z^+}$ so that $\theta-s+t$, $\sqrt{(\theta-s+t)^2+4n_2}$ and $\sqrt{4n_2(n_1-1)^2}$ are all integer multiples of $\sqrt{\Delta}$ for any signless Laplacian eigenvalue $\theta\in\mathrm{supp}_G(v)\setminus\{2r_1\}$. Furthermore, $\Delta$ divides $n_2$ at this point.
\end{enumerate}

\begin{proof}
As a result of Theorem 3.1, the eigenvalue support of vertex $(v,0)$ in $Q(G\tilde{\circ}{H})$ has the following form:
\begin{equation*}
\mathrm{supp}_{G\tilde{\circ}H}(v,0)=\{\theta_{\pm}:\theta\in \mathrm{supp}_G(v)\}.
 \end{equation*}
Notice that $r_{\pm}$ always belong to $\mathrm{supp}_{G\tilde{\circ}H}(v,0)$.

First we prove (i). Let $2r_1+t\neq s$. Assume that  $\theta-s+t$, $\sqrt{(\theta-s+t)^2+4n_2}$ and $\sqrt{(2r_1-s+t)^2+4n_2(n_1-1)^2}$ are all integers for any $\theta\in\mathrm{supp}_G(v)\setminus\{2r_1\}$. It is clear that vertex $(v,0)$ is periodic in $G\tilde{\circ}{H}$ according to Theorem 2.2. Conversely, suppose $(v,0)$ is a periodic vertex in $G\tilde{\circ}{H}$. Next we will divide the elements in $\mathrm{supp}_G(v)$ into two cases as follows:
\begin{enumerate}[(a)]
\item Each $\theta\in \mathrm{supp}_G(v)$ is integer. At this point, we find
\begin{equation*}
\theta_++\theta_--2s=\theta-s+t,
 \end{equation*}
 \begin{equation*}
\theta_+-\theta_-=\sqrt{(\theta-s+t)^2+4n_2}
 \end{equation*}
and
 \begin{equation*}
 r_+-r_-=\sqrt{(2r_1-s+t)^2+4n_2(n_1-1)^2}.
 \end{equation*}
These conclude that the aforementioned equations also have integers on their right-hand side..

\item Each $\theta\in \mathrm{supp}_G(v)$ is quadratic integer, and has the form of $\theta_\pm=\frac12(a+b_{\theta\pm}\sqrt{\Delta})$, where $a,b_{\theta_\pm}\in \mathbb{Z}$ and $\Delta\neq 1$ is a square-free integer. It follows from (\ref{7}) that $\prod_\pm(\theta_\pm-s)=-n_2$ for $\theta\in\mathrm{supp}_G(v)\setminus\{2r_1\}$ and $\prod_\pm(r_\pm-s)=-n_2(n_1-1)^2$. Combining these two equations, we have
\begin{equation*}
-n_2=\frac{1}{4}\left((a-2s)^2+b_{\theta_+}b_{\theta_-}\Delta\right)+\frac{1}{4}(a-2s)(b_{\theta_+}+b_{\theta_-})\sqrt{\Delta},
 \end{equation*}
 \begin{equation*}
-n_2(n_1-1)^{2}=\frac14\left((a-2s)^2+b_{r_+}b_{r_-}\Delta\right)+\frac14(a-2s)(b_{r_+}+b_{r_-})\sqrt{\Delta}.
 \end{equation*}
Since $\sqrt{\Delta}$ is an irrational number, thus either $b_{\theta_+}+b_{\theta_-}=0$ or $a-2s=0$. First assume that $b_{\theta_+}+b_{\theta_-}=0$. At that time, one has
 \begin{equation*}
a=\theta_++\theta_-=\theta+a-2s=r_{+}+r_{-}=r+a-2s,
 \end{equation*}
 which means $|\mathrm{supp}_{G}(v)|=1$, a contradiction. So, there must be $a-2s=0$. Then from $\theta_{\pm}=s+\frac{1}{2}b_{\theta_{\pm}}\sqrt{\Delta}$, one can obtain
 \begin{equation*}
\frac{1}{2}(b_{r_+}+b_{r_-})\sqrt{\Delta}=(r_+-s)+(r_--s)=(2r_1+t)-s\neq 0
 \end{equation*}
which is impossible because the left side of the equation above is irrational.
 \end{enumerate}
This concludes the proof of (i).

In what follows, we will prove (ii). Let $2r_1+t=s$. Assume that, for any signless Laplacian eigenvalue $\theta\in\mathrm{supp}_G(v)\setminus\{2r_1\}$, there exists a square-free $\Delta\in\mathbb{Z^+}$($\Delta\neq 1$) such that $\theta-s+t$, $\sqrt{(\theta-s+t)^2+4n_2}$ and $\sqrt{4n_2(n_1-1)^2}$ are all integer multiples of $\sqrt{\Delta}$. Then, it follows from Theorem 2.2 that $(v,0)$ is periodic in $G\tilde{\circ}{H}$.
Regarding the necessity, using the same approach above, we can get the following two cases: $b_{\theta_+}+b_{\theta_-}=0$ and $a-2s=0$. Exclude the first case as before. And if $a-2s=0$, then from $\theta_{\pm}=s+\frac{1}{2}b_{\theta_{\pm}}\sqrt{\Delta}$, one can obtain
\begin{equation*}
\frac{1}{2}(b_{\theta_+}+b_{\theta_-})\sqrt{\Delta}=\theta_++\theta_--2s=\theta-s+t,
 \end{equation*}
 \begin{equation*}
\frac{1}{2}(b_{\theta_+}-b_{\theta_-})\sqrt{\Delta}=\theta_+-\theta_-=\sqrt{(\theta-s+t))^{2}+4n_{2}},
 \end{equation*}
and
 \begin{equation*}
\frac{1}{2}(b_{r_+}-b_{r_-})\sqrt{\Delta}=r_+-r_-=\sqrt{4n_2(n_1-1)^2}.
 \end{equation*}
From the three equations, we can find that the right side of the equations are multiples of $\sqrt{\Delta}$ by half-integers. Their squares must be integers because they are all algebraic integers. Consequently, the right side of the equations are all integer multiples of $\sqrt{\Delta}$. Finally,
since $\sqrt{4n_2(n_1-1)^2}$ is integer multiples of $\sqrt{\Delta}$, we can easily conclude that $\Delta\mid n_2$.
\end{proof}

\paragraph{Theorem 4.3.} Suppose that the graphs $G$ and $H$ are stated as in Theorem 3.1. If $(v,0)$ is periodic in $G\tilde{\circ}{H}$, then
 \begin{equation*}
n_2\ge|\theta-s+t|+1, \quad\forall\;\theta\in\mathrm{supp}_G(v)\setminus\{2r_1\}
 \end{equation*}
and
 \begin{equation*}
n_2(n_1-1)^2\geq|2r_1-s+t|+1.
 \end{equation*}

\begin{proof} First assume that $2r_1+t\neq s$. Since $(v,0)$ is a periodic vertex in $G\tilde{\circ}{H}$. Thus, Theorem 4.2(i) states that $\theta-s+t$, $\sqrt{(\theta-s+t)^2+4n_2}$ and $\sqrt{(2r_1-s+t)^2+4n_2(n_1-1)^2}$ are all integers for any signless Laplacian eigenvalue $\theta\in\mathrm{supp}_G(v)\setminus\{2r_1\}$. It is apparent that $4n_2$ and $4n_2(n_1-1)^2$ are even number, the parities of $\theta-s+t$ and $\sqrt{(\theta-s+t)^2+4n_2}$ are same. It follows that
\begin{equation*}
\begin{aligned}
4n_2\ge\left(|\theta-s+t|+2\right)^2-|\theta-s+t|^2=4\left(|\theta-s+t|+1\right),\quad\forall\;\theta\in\operatorname{supp}_G(v)\setminus\{2r_1\}.
\end{aligned}
\end{equation*}
As for $\theta=2r_1$, since the parities of $2r_1-s+t$ and $\sqrt{(2r_1-s+t)^2+4n_2(n_1-1)^2}$ are also the same, then
\begin{equation*}
4n_2(n_1-1)^2\geq\left(|2r_1-s+t|+2\right)^2-\left(|2r_1-s+t|\right)^2=4\left(|2r_1-s+t|+1\right).
\end{equation*}
Hence, the conclusion follows.

Next, assume that $2r_1+t=s$. At that time, the inequality $n_2(n_1-1)^2\geq1$ is trivial. In this way we only need to prove that the first inequality holds. Theorem 4.2(ii) asserts that, for any signless Laplacian eigenvalue $\theta\in\mathrm{supp}_G(v)\setminus\{2r_1\}$, $\theta-s+t$, $\sqrt{(\theta-s+t)^2+4n_2}$ and $\sqrt{4n_2(n_1-1)^2}$ are all integer multiples of $\sqrt{\Delta}$. It is clear that the squares of $(\theta-s+t)/\sqrt{\Delta}$ and $\sqrt{(\theta-s+t)^2+4n_2})/\sqrt{\Delta}$ also have the same parity. Notice that $\Delta$ divides $n_2$ from Theorem 4.2(ii). Consequently, one obtains
\begin{equation*}
\frac{4n_2}\Delta\geq\left(\frac{|\theta-s+t|}{\sqrt{\Delta}}+2\right)^2-\left(\frac{|\theta-s+t|}{\sqrt{\Delta}}\right)^2=4\left(\frac{|\theta-s+t|}{\sqrt{\Delta}}+1\right),\quad\forall\;\theta\in\operatorname{supp}_G(v)\setminus\{2r_1\}.
\end{equation*}
Upon reorganizing, we conclude that the first inequality still holds, which finishes our proof.
\end{proof}

\paragraph{Example 1.} Let $G=C_4$ and $H$ be any $r_2$-regular graph. Noting that, at this time,  $t=3n_2$ and $s=2r_2+3$. Now, for any $v\in V(G)$ and signless Laplacian eigenvalue $\theta\in \text {supp}_G(v)\setminus\{2r_1\}$, we have $n_2\ge|\theta+n_2+2(n_2-r_2-1)-1|+1>n_2$, which is impossible. Hence, $(v,0)$ is not periodic vertex from Theorem 4.3. Consequently, $C_4\tilde{\circ}{H}$ has no signless-Laplacian-PST from Theorem 2.1.\\

As stated in the previous paragraph, a graph $G$ avoids signless-Laplacian-PST when $G$ has no periodic vertices. Next, we shall give some sufficient conditions for $G\tilde{\circ}{H}$ to lack periodic vertices.

\paragraph{Theorem 4.4.} Suppose $G$ and $H$ are described as in Theorem 3.1. Let $v\in V(G)$, for some square-free $\Delta\in\mathbb{Z}$, if either of the subsequent conditions holds:
 \begin{enumerate}[(i)]
 \item There exist two distinct elements $\lambda,\mu\in \mathrm{supp}_G(v)\setminus\{2r_1\}$, so that
  \begin{equation*}
  |\lambda-s+t|-|\mu-s+t|\in\left\{\sqrt{\Delta},2\sqrt{\Delta}\right\};
   \end{equation*}
 \item There exists an element $\gamma\in \mathrm{supp}_G(v)\setminus\{2r_1\}$, so that
  \begin{equation*}
\Bigl||2r_1-s+t|-(n_1-1)|\gamma-s+t|\Bigr|\in\left\{\sqrt{\Delta},2\sqrt{\Delta}\right\}.
   \end{equation*}
\end{enumerate}
Then $(v,w)$ is nonperiodic in $G\tilde{\circ}{H}$ for any $w\in V(H)\cup\{0\}$.

\begin{proof} First we prove (i). We just need to demonstrate that $(v,0)$ is nonperiodic on $G\tilde{\circ}{H}$ based on Lemma 4.1. Suppose by contradiction that $(v,0)$ is periodic in $G\tilde{\circ}{H}$. Thus, Theorem 4.2 implies that $\theta+s+t$ and $\sqrt{(\theta-s+t)^2+4n_2}$ are all integer multiples of $\sqrt{\Delta}$ for any signless Laplacian eigenvalue $\theta\in \mathrm{supp}_G(v)\setminus\{2r_1\}$. Set
\begin{equation*}
\delta=\frac{1}{\sqrt{\Delta}}\min\left\{\Bigl||\theta_1-s+t|-|\theta_2-s+t|\Bigr|:\theta_1,\theta_2\in\mathrm{supp}_G(v)\setminus\{2r_1\}\right\}.
 \end{equation*}
Suppose $\lambda$ and $\mu$ are the two different signless Laplacian eigenvalues contained in $\mathrm{supp}_G(v)\setminus\{2r_1\}$ which make the above equation minimum. Take
 \begin{equation*}
 n_\lambda=\frac{|\lambda-s+t|}{\sqrt{\Delta}},\quad n_\mu=\frac{|\mu-s+t|}{\sqrt{\Delta}}.
  \end{equation*}
Thus $\delta = n_{\lambda}-n_{\mu}$ and $\delta = 1,2$ from the hypothesis of theorem. Also Let
 \begin{equation*}
 z_1=\sqrt{n_\mu^2+\frac{4n_2}{\Delta}},\quad z_2=\sqrt{n_\lambda^2+\frac{4n_2}\Delta},
 \end{equation*}
 then $z_1,z_2\in \mathbb{Z}$ and
 $z_1+z_2>n_{\lambda}+n_{\mu}=2n_{\mu}+\delta$ and $z_2^{2}-z_1^{2}=(2n_{\mu}+\delta)\delta$,
 which means $z_2-z_1<\delta$. It is clear that this is not true whenever $\delta = 1$. If $\delta = 2$, then we still get a contradiction by the fact that $z_1$ has the same parity as $z_2$.

To prove (ii), using the similar approach, assume $(v,0)$ is periodic. Thus, two scenarios arise as hereunder mentioned:
\begin{enumerate}[(a)]
\item $2r_1+t\neq s$. Set
\begin{equation*}
\sigma=\min\left\{\Bigl||2r_{1}-s+t|-(n_1-1)|\gamma-s+t|\Bigr|:\gamma\in\mathrm{supp}_{G}(v)\setminus\{2r_1\}\right\}.
 \end{equation*}
Suppose that the signless Laplacian eigenvalue $\lambda$ makes the above minimum, let
\begin{equation*}
n_r=|2r_{1}-s+t|,\quad n_\lambda=|\lambda-s+t|,
 \end{equation*}
thus $\sigma=|n_r-(n_1-1)n_\lambda|$ and $\sigma = 1,2$. Also let
 \begin{equation*}
 s=\sqrt{n_\lambda^2+4n_2},\quad t=\sqrt{n_r^2+4n_2(n_1-1)^2}.
  \end{equation*}
Thus, $(n_1-1)s+t>(n_1-1)n_\lambda+n_r$ and $|t^2-((n_1-1)s)^2|=((n_1-1)n_\lambda+n_r)\sigma$, which means $|t-(n_1-1)s|<\sigma $. Clearly, this doesn't hold when $\sigma = 1$. If $\sigma = 2$, we get a contradiction by the fact that $(n_1-1)s$ has the same parity as $t$.
\item $2r_1+t=s$. Set
\begin{equation*}
\sigma=\frac{1}{\sqrt{\Delta}}\min\left\{\Bigl|(n_1-1)|\gamma-t+2r_{1}|\Bigr|:\gamma\in\operatorname{supp}_G(v)\setminus\{2r_1\}\right\}.
 \end{equation*}
Let $\lambda$ be the signless Laplacian eigenvalue which makes the above minimum. For convenience, take
\begin{equation*}
n_\lambda=\frac{|\lambda-s+t|}{\sqrt{\Delta}},
 \end{equation*}
thus $\sigma=(n_1-1)n_\lambda $ and $\sigma = 1,2$. Let
 \begin{equation*}
 s=\sqrt{n_\lambda^2+\frac{4n_2}{\Delta}},\quad t=\sqrt{\frac{4n_2(n_1-1)^2}{\Delta}}.
  \end{equation*}
Then, $(n_1-1)s+t>\sigma$ and $|t^2-((n_1-1)s)^2|=\sigma^2$, which means $|t-(n_1-1)s|<\sigma $. Similar to (i), we can get a contradiction.
\end{enumerate}

To sum up, this finishes our proof.
 \end{proof}

\paragraph{Theorem 4.5.} Suppose the graphs $G$ and $H$ are described as in Theorem 3.1. Let $v\in V(G)$, if either of the subsequent conditions holds:
\begin{enumerate}[(i)]
\item There exist two distinct elements $\lambda, \mu\in\mathrm{supp}_G(v)\setminus\{2r_1\}$, so that
\begin{equation}\label{9}
0< |\lambda-s+t|-|\mu-s+t|<3;
\end{equation}
\item There exists an element $\gamma\in \mathrm{supp}_G(v)\setminus\{2r_1\}$, so that
\begin{equation}\label{10}
0< \Bigl||2r_1-s+t|-(n_1-1)|\gamma-s+t|\Bigr|<3.
\end{equation}
\end{enumerate}
Then, $(v,w)$ is nonperiodic on $G\tilde{\circ}{H}$ for any $w\in V(H)\cup\{0\}$.

\begin{proof}
(i) According to Lemma 4.1, it suffices to prove that $(v,0)$ is a nonperiodic vertex on $G\tilde{\circ}{H}$. Suppose towards contradiction that $(v,0)$ is periodic. From Theorem 4.2 and (\ref{9}), one obtains
\begin{equation*}
|\lambda-s+t|-|\mu-s+t|\in\{\sqrt{1},\sqrt{2},\sqrt{3},2\sqrt{1},\sqrt{5},\sqrt{6},\sqrt{7},2\sqrt{2}\},
\end{equation*}
which implies that $(v,0)$ is not periodic using Theorem 4.4 (i).

(ii) Suppose that $(v,0)$ is periodic. Use the same way as above, it follows from Theorem 4.2 and (\ref{10}) that
\begin{enumerate}[(a)]
\item If $2r_1+t\neq s$, then $\begin{vmatrix}|2r_1-s+t|-(n_1-1)|\gamma-s+t|\end{vmatrix}\in\{\sqrt{1},2\sqrt{1}\}$;
\item If $2r_1+t= s$, then $\begin{vmatrix}(n_1-1)|\kappa-s+t|\end{vmatrix}\in\{\sqrt{1},\sqrt{2},\sqrt{3},2\sqrt{1},\sqrt{5},\sqrt{6},\sqrt{7},2\sqrt{2}\}$.
\end{enumerate}
Hence, it follows from Theorem 4.4 (ii) that $(v,0)$ is periodic on $G\tilde{\circ}{H}$, a contradiction.
\end{proof}

\paragraph{Example 2.} We state that $G\tilde{\circ}{H}$ lacks signless-Laplacian-PST if $H$ is $K_1$ and $G$ is a graph of which has one of the form as below:
\begin{enumerate}[(i)]
\item Hamming $d$-dimensional cubes with $2^d$ vertices;
\item Halved $2d$-dimensional cubes with $2^{2d-1}$ vertices;
\item Double coset graph of binary Golay code with 4096 vertices.
\end{enumerate}
\begin{proof}
Note that the graph families above are distance-regular, implying that for each vertex in $G$, its signless Laplacian eigenvalue support includes all the signless Laplacian eigenvalues of $G$ (see \cite{Coutinho14,Coutinho15}). From \cite{Coutinho15}, we can find that all the distinct spectral of $Q(G)$ is (i) $\{2d-2k:k=0,1,\ldots,d\}$; (ii) $\left\{2\binom{2d}{2}-2k(2d-k):k=0,1,\ldots,d\right\}$; (iii) $\{46, 32, 30, 24, 22, 16, 14, 0\}$, respectively. By simple calculation, all the three sets of signless Laplacian eigenvalues conform to the inequalities in Theorem 4.5. Thus, all the vertices in  $G\tilde{\circ}{H}$ are not periodic. Therefore, $G\tilde{\circ}{K_1}$ avoids signless-Laplacian-PST from Theorem 2.1.
\end{proof}

It is observed that $K_2\circ{H}$ and $K_2\tilde{\circ}{H}$ are the same. Tian et al. \cite {Tian21} showed that $K_2\tilde{\circ}{H}$ lacks signless-Laplacian-PST between two vertices of $K_2$ when $n_2$ is even number. Following this, we find that $K_2\tilde{\circ}{H}$ still lacks signless-Laplacian-PST between the two inner vertices when $n_2$ is a prime number or 1.

\paragraph{Theorem 4.6.} Let $G=K_2$ with vertices $u$ and $v$, and $H$ be an $r_2$-regular graph with $n_2$ vertices. If $n_2$ is a prime number or 1, then $K_2\tilde{\circ}{H}$ avoids signless-Laplacian-PST.

\begin{proof}
First noting that if $G=K_2$, there are only two signless Laplacian eigenvalues 0 and 2 in $G$. At this point, it is clear that $\theta_\pm$ and $r_\pm$ have the same expression. Now, from Theorem 4.2, we find that $K_2\tilde{\circ}{H}$ has periodic vertex $(v,0)$ if and only if $2n_2$ is divided by a square-free $\Delta$ in which case $\theta-s+t$ and $\sqrt{(\theta-s+t)^2+4n_2}$ are integer multiples of $\sqrt{\Delta}$ for $\theta\in \{0,2\}$.

Since the periodicity is a requirement for two vertices to allow signless-Laplacian-PST. Thus we prove the conclusion by contradiction. Assume that $(v,0)$ is periodic in $K_2\tilde{\circ}{H}$, noting that $s=2r_2+1$, $t=n_2$, then $\theta+n_2-2r_2-1$ and $\sqrt{(\theta+n_2-2r_2-1)^2+4n_2}$ are integer multiples of $\sqrt{\Delta}$ for $\theta\in \{0,2\}$. Set
\begin{equation*}
m_0=\frac{n_2-2r_2-1}{\sqrt{\Delta}},\quad m_2=\frac{2+n_2-2r_2-1}{\sqrt{\Delta}},\quad l=\frac{2n_2}\Delta,
\end{equation*}
and $M_0^2=m_0^2+2l$, $M_2^2=m_2^2+2l$ for two positive numbers $M_0, M_2$. Obviously, $M_0, M_2\in \mathbb{Z}$.

First assume that $n_2$ equals to 1, then $l=1$ or $2$. If $l=1$, thus $M_0^2-m_0^2=2$, but it is impossible that the difference between two squares equals to 2. If $l=2$, then $M_0^2-m_0^2=4$ and $M_2^2-m_2^2=4$, only if $M_0=M_2=2$ and $m_0=m_2=0$ the equation holds, a contradiction.

If $n_2$ equals to 2, then $l=2$ or $4$. We may exclude the first possibility based on the proof before. As for $l=4$, since $m_0\neq m_2$, there is exactly one condition in which $M_0=M_2=3$, $m_0=-1$ and $m_2=1$. In this case, one can get $n_2=2, r_2=1$, viz. $H=K_2$. According to Theorem 4.7 in \cite{Tian21}, there is no signless-Laplacian-PST in $K_2\tilde{\circ}{K_2}$.

If $n_2$ is an odd prime, then $l=1, 2, n_2$ or $2n_2$. In the same way as the proof before, we may exclude the first two possibilities. If $l=n_2$, then $M_0^2-m_0^2=2n_2$ is an even number. Since $n_2=(M_0-m_0)((M_0+m_0)/2)$ gives us $M_{0}-m_{0}=1$, a contradiction. If $l=2n_2$, then $n_2=((M_0-m_0)/2)((M_0+m_0)/2)=((M_2-m_2)/2)((M_2+m_2)/2)$ gives us $M_0-m_0=M_2-m_2=2$. In this case, $n_2=m_{0}+1=m_{2}+1$, a contradiction.

To sum up, $(v,0)$ is nonperiodic on $K_2\tilde{\circ}{H}$. It follows from Lemma 4.1 that $K_2\tilde{\circ}{H}$ lacks periodic vertices. Hence, $K_2\tilde{\circ}{H}$ avoids signless-Laplacian-PST.
\end{proof}

Now we have known that $K_2\tilde{\circ}{H}$ lacks signless-Laplacian-PST in two cases where $n_2$ is prime number or 1. Together with Theorem 4.7 in \cite {Tian21} for the case where $n_2$ is even number, we can reasonably suspect that $K_2\tilde{\circ}{H}$ avoids signless-Laplacian-PST for arbitrary $n_2$.

\section{Signless-Laplacian-PGST}

In the prior chapter, we found that the signless-Laplacian-PST in $G\tilde{\circ}H$ is rare. Now we shall focus on signless-Laplacian-PGST in the following part and provide some vertex complemented coronae to allow signless-Laplacian-PGST under suitable conditions.

\paragraph{Theorem 5.1.} Suppose $G$ is an $r_1$-regular connected graphs with  $n_1\geq2$ vertices and allows signless-Laplacian-PST from the vertex $u$ to the vertex $v$ in $G$ at time $\tau=\pi/g$ with $g\in\mathbb{Z^+}$. Let $\Lambda_r$ be described as in Theorem 3.2. If $\Lambda_r$ is irrational, then $G\tilde{\circ}\overline{{K_{n_2}}}$ allows signless-Laplacian-PGST from the vertex $(u,0)$ to the vertex $(v,0)$.

\begin{proof} Firstly, when $n_1=2$, it follows from Theorem 5.2 in \cite{Tian21} that $K_2\tilde{\circ}\overline{{K_{n_2}}}$ allows signless-Laplacian-PGST from $(u,0)$ to $(v,0)$. After that, we merely need to consider the case when $n_1\geq3$. Since $G$ allows signless-Laplacian-PST at time $\tau=\pi/g$, using Theorem 2.2, $\Delta=1$. As we can see from Theorem 2.2 again, any signless Laplacian eigenvalue $\theta\in \mathrm{supp}_G(v)$ is an integer. Next, we assert that each $\Lambda_{\theta}$ is irrational, in which $\Lambda_{\theta}$ be described as in Theorem 3.2. Indeed, assume that there exists some $\theta_*\in \mathrm{supp}_G(v)\setminus\{2r_1\}$ such that $\Lambda_{\theta_*}$ is an integer. Since the parity of $\theta_*-s+t$ and $\sqrt{(\theta_*-s+t)^2+4n_2}$ are same, implying that  $4n_2\ge\left(|\theta_*-s+t|+2\right)^2-|\theta_*-s+t|^2$. By simplification, we have $n_2\ge|\theta_*-s+t|+1$. Noting that $t=n_2(n_1-1)$ and $s=n_1-1$, one obtains that $\theta_*\leq(n_2-1)(2-n_1)<0$ for $n_1\geq3$, a contradiction.  Thus, there must exist integer $s_{\theta}$ and $s_{r}$ so that $\Lambda_\theta=s_\theta\sqrt{c_\theta}$ and $\Lambda_r=s_r\sqrt{c_r}$, where $c_\theta$ and $c_r$ are all square-free. Notice that $2r_1\in \mathrm{supp}_G(v)$, it follows from Theorem 2.4 that,
\begin{equation*}
\{1\}\cup\{\sqrt{c_\theta}:\;\theta\in\mathrm{supp}_G(v)\}
\end{equation*}
is linearly independent over $\mathbb{Q}$. On the basis of Theorem 2.3, we arrive at
\begin{equation*}
l\sqrt{c_\theta}-q_\theta\approx-\frac{\sqrt{c_\theta}}{2g},\quad\forall\;\theta\in\mathrm{supp}_G(v)
\end{equation*}
for some integers $l$ and $q_\theta$. Now, $4s_\theta$ is multiplied on both sides of the equationa bove, one obtains
\begin{equation}\label{11}
\left(4l+\frac2g\right)\Lambda_\theta\approx 4q_\theta s_\theta,\quad\forall\;\theta\in\mathrm{supp}_G(v).
\end{equation}
Notice that the case $\theta=2r_1$ is also included in (\ref{11}). Set $T=(4l+\frac{2}{g})\pi$, it is easy to see that $\cos(\Lambda_\theta \frac{T}{2})\approx1$ and $\sin(\Lambda_\theta \frac{T}{2})\approx0$. Keep in mind that $|\mathrm{e}^{-\mathrm{i}T\frac{s+t}{2}}|$ always equals to 1. Now, from Theorem 3.2, we derive
\begin{equation*}
\begin{aligned}
&\left|\mathbf{e}_{(u,0)}^{T} \mathrm{e}^{-\mathrm{i}TQ(G\tilde{\circ}\overline{{K_{n_2}}})}\mathbf{e}_{(v,0)} \right| \\
&\quad=\left| \mathrm{e}^{-\mathrm{i}T\frac{s+t}{2}} \right|\bigg| \sum_{\theta\in\mathrm{Spec}_{Q}(G)\setminus\{2r_1\}}\mathrm{e}^{-\mathrm{i}T\frac{\theta}{2}}\mathbf{e}_u^T F_\theta(G)\mathbf{e}_v\left(\cos\left(\frac{\Lambda_{\theta}T}{2}\right)-\mathrm{i}\frac{\theta-s+t}{\Lambda_{\theta}}\sin\left(\frac{\Lambda_{\theta}T}{2}\right)\right) \\
&\quad\quad+ \text e^{-\mathrm{i}T\frac{2r_1}{2}}\mathbf{e}_u^T F_{2r_1}(G)\mathbf{e}_v\left(\cos\left(\frac{\Lambda_rT}{2}\right)-\mathrm{i}\frac{2r_1-s+t}{\Lambda_r}\sin\left(\frac{\Lambda_rT}{2}\right)\right) \bigg| \\
&\quad\approx \left| \sum_{\theta\in\mathrm{Spec}_{Q}(G)}\mathrm{e}^{-\mathrm{i}(2\pi)l\theta}\mathrm{e}^{-\mathrm{i}(\frac{\pi}{g})\theta}\mathbf{e}_u^T F_{\theta}(G)\mathbf{e}_{v} \right| \\
&\quad=\left| \mathbf{e}_u^T\mathrm{e}^{-\mathrm{i}(\frac{\pi}{g})Q(G)}\mathbf{e}_v \right|.
\end{aligned}
\end{equation*}
Finally, since $G$ allows signless-Laplacian-PST from the vertex $u$ to the vertex $v$ at $\tau=\pi/g$, hence the desired result holds.
\end{proof}

\paragraph{Example 3.} Given a coset graph $G$ of which is the shortened binary Golay code, it is well known \cite{Coutinho15} that $n_1=2048, r_1=22$ and the set of all the distinct signless Laplacian eigenvalues of $G$ has the form $\{44, 30, 28, 22, 20, 14, 12\}$. Let $H=\overline{K_1}$, viz. $n_2=1$ and $r_2=0$. Since $G$ has signless-Laplacian-PST at time $\pi/2$ and $\Lambda_r=\sqrt{(44)^2+4\times2047^2}=\sqrt{16762772}$ is irrational. Therefore, according to Theorem 5.1, this graph $G\tilde{\circ}\overline{K_{1}}$ exhibits signless-Laplacian-PGST.\\

Remark that the cocktail party graph $\overline{mK_{2}}$ exhibits signless-Laplacian-PST at time $\pi/2$ only if $m$ is even (see \cite{Coutinho14}). At this moment, if $\Lambda_r=2\sqrt{4(m-1)^2+(2m-1)^2}$ is irrational, then $\overline{mK_{2}}\tilde{\circ}{K_1}$ exhibits signless-Laplacian-PGST from Theorem 5.1. What below demonstrates that the requirements of Theorem 5.1 can be relaxed for the cocktail party graph $\overline{mK_{2}}$ with odd number $m>2$. Additionly, these results also show that graph $\overline{mK_{2}}\tilde{\circ}{K_1}$ exhibits signless-Laplacian-PGST for almost all cocktail party graphs $\overline{mK_{2}}$.

\paragraph{Theorem 5.2.} Suppose $G$ is the cocktail party graph $\overline{mK_{2}}$ with odd number $m>2$, and $u, v$ are its two antipodal vertices. If any of the following conditions exists:
\begin{enumerate}[(i)]
	\item $\sqrt{4(m-1)^2+(2m-1)^2}$ is an integer;
	\item $\sqrt{4(m-1)^2+(2m-1)^2}$ is irrational, $\sqrt{4(m-1)^2+(2m-1)^2}$ and $\sqrt{(m-1)^2+1}$ have the different square-free parts.
\end{enumerate}
Then $\overline{mK_{2}}\tilde{\circ}{K_1}$ exhibits signless-Laplacian-PGST from $(u,0)$ to $(v,0)$.

\begin{proof}
It is known that the distinct signless Laplacian eigenvalues of $\overline{mK_{2}}$ are $\theta_0=2r_1=4(m-1)$, $\theta_1=2(m-1)$ and $\theta_2=2(m-2)$. Assume their respective eigenprojectors $F_0$, $F_1$ and $F_2$. From this, we have $\Lambda_r=2\sqrt{4(m-1)^2+(2m-1)^2}$, $\Lambda_{\theta_1}=2\sqrt{(m-1)^2+1}$ and $\Lambda_{\theta_2}=2\sqrt{(m-2)^2+1}$. It is easy to verify that $\Lambda_{\theta_1}$ and $\Lambda_{\theta_2}$ are irrational with different square-free parts for $m>2$. As a matter of convenience, let $\Lambda_{\theta_i}=2s_{\theta_i}\sqrt{c_{\theta_i}}$ where $s_{\theta_i}$ is integer and $c_{\theta_i}$ is square-free for $i=1,2$.

To prove (i), assume that $\sqrt{4(m-1)^2+(2m-1)^2}$ is an integer. It is clearly shown in Theorem 2.3 that there must exist $l, q_1, q_2 \in \mathbb{Z}$ so that
\begin{equation*}
l\sqrt{c_{\theta_1}}-q_1\approx\frac12,\quad l\sqrt{c_{\theta_2}}-q_2\approx0.
\end{equation*}
Observe that $s_{\theta_1}$ is odd integer as $m$ is odd.
Therefore, if we choose $T=2\pi l$, then $\cos(\Lambda_{r}T/2)=1$ and $\cos(\Lambda_{\theta_i}T/2)\approx(-1)^i$ for $i=1, 2$. And at the same time, $\mathrm{e}^{-\text{i}T((\theta_i+s+t)/2)}=1$ for $i=0, 1, 2$.
In addition, if follows from Theorem 2.5 that
\begin{equation*}
\mathbf{e}_u^T F_i(\overline{mK_2})\mathbf{e}_v=(-1)^i\mathbf{e}_u^T F_i(\overline{mK_2})\mathbf{e}_u.
\end{equation*}
Hence, $\mathbf{e}_{(u,0)}^T\mathrm{e}^{-\text{i}TQ(\overline{mK_2}\tilde{\circ}{K_1})}\mathbf{e}_{(v,0)}\approx1$, implying the expected result.

In what follows, we prove (ii). In this case, since $\sqrt{4(m-1)^2+(2m-1)^2}$ is irrational, then there must exist integer $s_{\theta_0}$ and square-free $c_{\theta_0}$ such that $\Lambda_{r}=2s_{\theta_0}\sqrt{c_{\theta_0}}$. It follows from Theorem 2.4 that
\begin{equation*}
\{1\}\cup\{\sqrt{c_{\theta_i}}:\;{\theta_i}\in\mathrm{supp}_G(v)\}
\end{equation*}
is linearly independent over $\mathbb{Q}$. Since $\sqrt{4(m-1)^2+(2m-1)^2}$ and $\sqrt{(m-1)^2+1}$ have the different square-free parts, then $c_{\theta_0}\neq c_{\theta_1}$. Thus, by Theorem 2.3 again, there must exist $l, q_0, q_1, q_2 \in \mathbb{Z}$ so that
\begin{equation*}
l\Lambda_{\theta_0}-q_0\approx0,\quad\ \\l\Lambda_{\theta_1}-q_1\approx\frac12,\quad\ \\l\Lambda_{\theta_2}-q_2\approx0.
\end{equation*}
The rest of the process is analogous to (i), and the details are ignored.
\end{proof}
\noindent
\textbf{Acknowledgements}
This work was in part supported by the National Natural Science Foundation of China (Nos. 11801521, 12071048).

\end{document}